\documentclass[11pt,preprintnumbers,aps,amssymb,nofootinbib,amsmath,superscriptaddress]
{revtex4}
\usepackage{epsfig,epsf}
\usepackage{bm} 
\usepackage{color}
\newcommand{\beq}{\begin{equation}}
\newcommand{\beql}[1]{\begin{equation}\label{#1}}
\newcommand{\eeq}{\end{equation}}
\newcommand{\eq}[1]{(\ref{#1})}
\newcommand{\fig}[1]{Fig.~\ref{#1}}
\renewcommand{\sec}[1]{Sec.~\ref{#1}}
\newcounter{topiccounter}
\renewcommand{\b}[1]{{\bm #1}} 

\newcommand{\as}{ \alpha_s}

\newcommand{\Tr}{{\rm Tr}}

\newcommand{\real}{\mathrm{Re}\,}

\newcommand{\aver}[1]{\left\langle #1 \right\rangle}

\begin{document}

\title{Beyond the proton collinear factorization in heavy quark production in pA collisions at low $x$.}

\author{Kirill Tuchin}

\affiliation{
Department of Physics and Astronomy, Iowa State University, Ames, IA 50011}

\date{\today}


\begin{abstract}

We consider heavy quark production in high energy pA collisions and investigate the contribution of  interactions of valence quarks of proton with the nucleus.  The often made assumption  that valence quarks of proton can be factored out is justified only if the nucleus saturation momentum is much smaller than the heavy quark mass.  This is not the case in phenomenologically relevant situations. Breakdown of factorization manifests itself in substantial decrease of the cross section at large total and small relative transverse momenta of the heavy quark -- antiquark pair.

\end{abstract}

\maketitle

\section{Introduction}\label{sec:intr}

The most general expression for heavy quark pair production in high energy pA collisions was derived in \cite{Kovchegov:2006qn}. It takes into account interaction of valence quark $q_v$, intermediate virtual gluon and the produced $q\bar q$ pair with the heavy nucleus target. That expression is quite bulky and unfriendly for numerical calculation because it involves multi-dimensional  nested integrals over the oscillating integrands. Therefore, one usually considers an approximation in which interaction of valence quark with the nucleus is neglected leading to collinear factorization of the gluon distribution function on the proton side \cite{Kopeliovich:2002yv,Tuchin:2004rb,Blaizot:2004wv,Kharzeev:2008cv,Kharzeev:2008nw,Dominguez:2011wm}. This approximation turns out to be valid only in the limit of very heavy quarks $m\gg Q_s$, where $Q_s$ is the saturation momentum of \emph{the target nucleus}.  Since, $Q_s$ turns out to be of the same order of magnitude as the charm and bottom quark masses, interactions of valence quarks must be taken into account. 
It is the goal of this article to develop an approximation that takes 
into account interaction of valence quark with the nucleus and still allows one to make relatively simple numerical analysis of heavy quark production. 

To develop a consistent approximation we need to ensure the gauge invariance of each term in the expansion. The derivation is presented in the main part of this article, which is structured as follows. In \sec{sec:vs} we review the result of  \cite{Kovchegov:2006qn}. In \sec{sec:limit} we discuss the factorization limit of these general formulas. Contribution of valence quarks is calculated in \eq{sec:ccc}. Our main result is given by Eqs.~\eq{main1},\eq{wp2},\eq{wx15}. The effect of  valence quark interactions on double inclusive cross section of charm production is shown in \fig{fig3}. We denoted the transverse momenta of quark and antiquark as $\b k_1$ and $\b k_2$,  the fraction of the gluon's light-cone energy carried by the heavy quark as $z$, the pair total momentum as $\b q= \b k_1+\b k_2$ and their relative momentum as $\b \ell = (1-z)\b k_1-z\b k_2$, which is related to the 
invariant mass of the pair as $M^2=(m^2+\b\ell^2)/z(1-z)$.  It is seen on this figure that deviation from the collinear factorization of proton is negligible  at small $q$, which corresponds to valence quark being collinear with the valence quark, and large $\ell$, which corresponds to large invariant mass of the pair. However, it grows as $q$ increases and $\ell$ decreases.  It appears that the collinear factorization of proton is a reasonable approximation for total cross sections which receive the main contribution from low transverse momenta. However, for single and double-inclusive spectra it overestimates the cross section by orders  of magnitude.

\section{Heavy quark and antiquark production in pA collisions: general result }\label{sec:vs}
\begin{figure}[ht]
      \includegraphics[height=4cm]{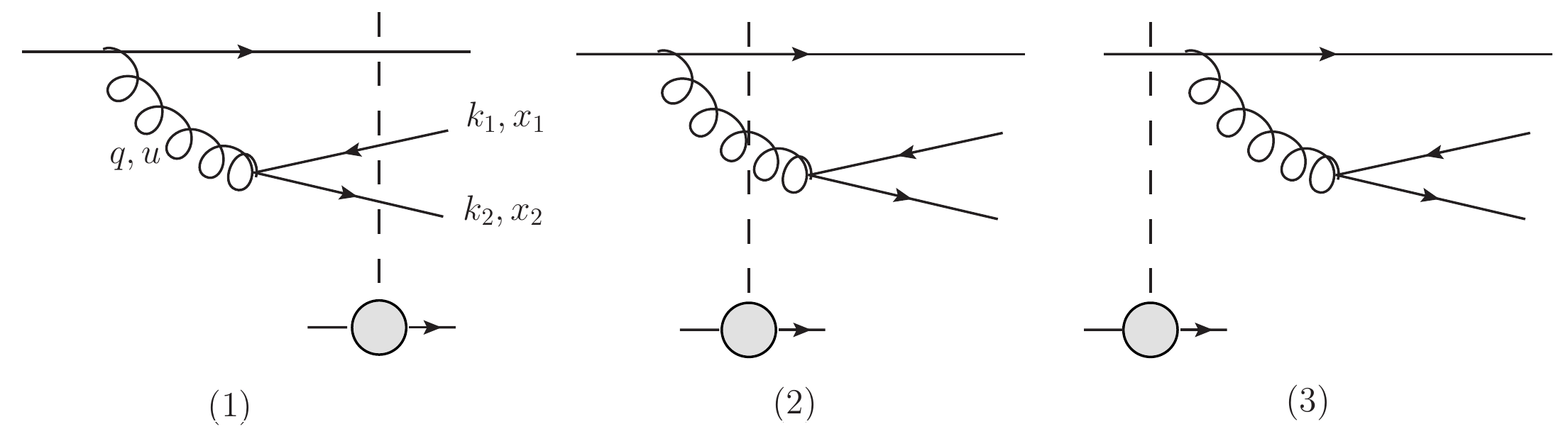} 
  \caption{Diagrams contributing to quark-antiquark pair production in the light-cone gauge. (1) Incoming valence quark emits  a gluon, which splits into a quark-antiquark pair before the system hits the target. (2) Valence quark first emits a gluon, after which the system
rescatters on the target nucleus, and later the gluon splits into a
quark-antiquark pair. (3) Valence quark rescatters on the target nucleus, after which it produces a gluon, which splits into a quark-antiquark pair.  }
\label{wave_funct}
\end{figure}

The diagrams contributing to quark-antiquark pair production in the light-cone gauge are shown in \fig{wave_funct}. The vertical dashed line depicts interaction of the projectile partonic system with the target. Since the production time of the partonic system is proportional to the collision energy, it is much larger than the interaction time, which is taken to be vanishingly small. The calculation is done in the light-cone perturbation theory 
\cite{Lepage:1980fj}, along the lines outlined in \cite{Kovchegov:1998bi}. It is convenient to work in coordinate space where the diagram contributions factorize into a convolution of Glauber-Mueller multiple rescattering \cite{Mueller:1989st} with the ``wave function'' parts, which include splittings $q_v \rightarrow q_v\, g$ and $g \rightarrow q\, \bar q$.
Since eikonal multiple rescatterings do not change the transverse coordinates of the incoming quarks and the gluon, we can calculate the ``wave functions"  in transverse coordinate space by calculating the diagrams in \fig{wave_funct} without interactions. We denote momenta of the outgoing quark and anti-quark as $k_1$ and $k_2$ correspondingly. We  assume that  the gluon is  much softer than the proton, i.e.\  $k_{1+} + k_{2+} \ll p_+$ $k_{1+}, k_{2+} \ll p_+$, where $p_+$ is the typical light cone momentum of the valence quarks in the proton. Expressions for the wave functions in the momentum space can be found in \cite{Kovchegov:2006qn}.
The light cone ``wave-functions'' in transverse coordinate space are defined as
\beq\label{mom.space}
\Psi^{(i)}_{\sigma,\, \sigma'} (\b x_1, \b x_2;  z ) \,=\, 
\int \frac{d^2 k_1}{(2\pi)^2} \frac{d^2 k_2}{(2\pi)^2} \, e^{-i\b k_1 
\cdot \b x_1 - i \b k_2 \cdot \b x_2} \, \Psi^{(i)}_{\sigma,\, \sigma'} 
(k_1,k_2)\,,\quad i=1,2,3\,,
\eeq
where the superscript $(i)$ corresponds to one of three diagrams in \fig{wave_funct}. Here we assume that the transverse coordinate of the valence quark, which emits the gluon (which splits into a $q\bar q$ pair) is $\b 0$ and denoted the transverse coordinate of quark and antiquark by $\b x_1$ and $\b x_2$ correspondingly. The quark-antiquark production cross section is proportional to the product of the sum of the light-cone ``wave-functions" in the amplitude and sum of the light-cone ``wave-functions" in the complex-conjugated (c.c.) one, averaged over the quantum numbers of the initial valence quark and summing over the quantum numbers of the final state quarks. Since transverse momenta of quark $\b k_1$ and antiquark $\b k_2$ are fixed, their coordinates in the amplitude and in the c.c.\  amplitude are different. We will denote the corresponding coordinates in the c.c.\ amplitude by $\b x_1'$ and $\b x_2'$. The resulting quantity appearing in the cross section is given by 
\beq\label{kernel}
\Phi_{ij}(\b x_1, \b x_2; \b x_1', \b x_2';  z ) \, = \, 
\frac{1}{N_c}\sum_{\sigma,\sigma', a,b} \, 
\Psi^{(i)}_{\sigma, \, \sigma'} (\b x_1, \b x_2;  z )
\, \Psi^{(j) *}_{\sigma, \, \sigma'} (\b x_1', \b x_2';  z ) \,,\quad i,j=1,2,3\,.
\eeq
Here the sum over gluons' colors $a$ and $b$ simply implies a
calculation of the color factors of the relevant diagrams, including
traces over fermion loops.

The  double-inclusive quark--anti-quark production cross section in $pA$ collisions in the
quasi-classical approximation, i.e.\ neglecting quantum evolution of the partonic system,
\begin{align}\label{dcl}
\frac{d \sigma^{q_v\to q\bar qX}}{d^2k_1 d^2 k_2 \, dy\,  d  z\,  d^2 b} = 
\frac{1}{4  (2  \pi)^6}  \int d^2 x_1 \, d^2 x_2 \, d^2 x'_1 \, d^2 x'_2 \, 
e^{- i \b k_1 \cdot (\b x_1 - \b x'_1) - i \b k_2 \cdot (\b x_2 - \b x'_2)} &\nonumber\\
\times \, \sum_{i,j=1}^3\, \Phi_{ij} (\b x_1, \b x_2; \b x_1', \b x_2';  z ) \
S_{ ij} (\b x_1, \b x_2; \b x_1', \b x_2';  z )&\,.
\end{align}
Here $y$ is the rapidity of the $s$-channel gluon, which splits into
the $q\bar q$ pair. $S_{ij}$ are the $S$-matrix elements corresponding to the $i$'th wave function in the amplitude and $j$'th in the complex-conjugated amplitude.   Since the quark and the anti-quark are most likely
to be produced close to each other in rapidity, one can think of $y$
as the rapidity of the quarks. $\b b$ is the impact parameter of the
proton with respect to the nucleus.

The single inclusive quark production cross section is obtained
from \eq{dcl} by integrating over one of the quark's momenta and multiplying by 2 to account for both quarks and anti-quarks:
\begin{align}\label{single_cl}
\frac{d \sigma^{q_v\to qX}}{d^2 k \, dy \, d^2 b} \,=\, \frac{1}{2 \, (2 \, \pi)^4} \, \int  
d^2 x_1 \, d^2 x_2 \, d^2 x_1'  \, \int_0^1 d z  \, e^{-i \, \b k \cdot (\b x_1-\b
x_1' )} \, & \nonumber\\
\times \,\sum_{i,j=1}^3 \, \Phi_{ij}\, (\b x_1, \b x_2; \b x_1' , \b
x_2;  z )\, S_{ij} (\b x_1, \b x_2; \b x_1' , \b x_2;  z )\,,&
\end{align}
where $y$ is the rapidity of the produced (anti-)quark.

Explicit expressions of the ``wave-functions" can be found in \cite{Kovchegov:2006qn}. Upon summation over $\lambda$ they read
\begin{align}
\Psi^{(1)}_{\sigma, \, \sigma'} (\b x_1, \b x_2;  z ) &=
\frac{2 \, g^2 \, T_a \, T_b}{(2 \pi)^2} \,
\bigg[ F_2 (\b x_1 ,\b x_2; z ) \, \frac{1}{x_{12} \, u} \, 
[ (1- 2 \,  z ) \, \b x_{12}\cdot \b u + i \, \sigma \, \epsilon_{ij} \,
u_i \, x_{12 \, j}] \, \delta_{\sigma,\sigma'}
\nonumber\\
&\,+ \, F_1 (\b x_1 ,\b x_2;  z ) \, \frac{i}{u} \, 
\sigma \, m \, (u_x + i \, \sigma \, u_y)
\, \delta_{\sigma, -\sigma'}  \,
\nonumber\\
&- 2 \, \delta_{\sigma, \, \sigma'} \,  z  \, (1 -  z ) \, F_0 (\b x_1 ,\b x_2;  z ) \, \bigg] ,\label{ms1}\\
\Psi^{(2)}_{\sigma,\, \sigma'} (\b x_1, \b x_2;  z ) &= 
- \frac{2 \, g^2 \, T_a \, T_b}{(2\pi)^2} \,
\bigg[ m \, K_1 (m\, x_{12}) \, \frac{1}{x_{12} \, u^2} \, 
[ (1- 2 \,  z ) \, \b x_{12}\cdot \b u + i \, \sigma \, \epsilon_{ij} \,
u_i \, x_{12 \, j}] \, \delta_{\sigma,\sigma'}
\nonumber\\
& + \, K_0(m \, x_{12}) \, \frac{i}{u^2} \, 
\sigma \, m \, (u_x + i \, \sigma \, u_y) \,
\delta_{\sigma,-\sigma'}\bigg] \,,\label{ms2}\\
\Psi^{(3)}_{\sigma, \, \sigma'} (\b x_1, \b x_2;  z ) &=
- \Psi^{(1)}_{\sigma, \, \sigma'} (\b x_1, \b x_2;  z ) 
- \Psi^{(2)}_{\sigma, \, \sigma'} (\b x_1, \b x_2;  z )\,,\label{ms3}
\end{align}
where $\epsilon_{12} = 1 = - \epsilon_{21}$, $\epsilon_{11} =
\epsilon_{22} = 0$, and, assuming summation over repeating indices, 
$\epsilon_{ij} \, u_i \, v_j = u_x \, v_y - u_y \, v_x$. Also $x_{12
\, j}$ denotes the $j$th component of the vector $\b x_{12}$. Transverse coordinates of gluon in the amplitude is
\beq\label{u}
\b u \, = \,  z  \, \b x_1 + (1- z ) \, \b x_2\,.
\eeq 
We denote $u = |\b u|$, $\b x_{12} = \b x_1 - \b x_2$, $x_{12} =
|\b x_{12}|$. Similarly, gluon's transverse coordinate in the c.c.\ amplitude is $\b
u' =   z   \b x_1' + (1 -  z )  \b x_2'$ with $u' =
|\b u'|$ and $\b x_{12}' = \b x_1' - \b x_2'$, $x_{12}' = |\b
x_{12}'|$.
To perform the Fourier transform of \eq{mom.space} we first 
introduce the following auxiliary functions
\begin{align}
F_2 (\b x_1, \b x_2; z)&= \int_0^\infty d q \, J_1 ( q u) \,
K_1 \bigg(x_{12} \, \sqrt{m^2+
 q^2\, z(1-z)} \bigg)\, \sqrt{m^2+
 q^2\, z(1-z)} \,,\label{aux.fun1}\\ 
F_1 (\b x_1, \b x_2;z)&= \int_0^\infty d q \, J_1( q  u) \, K_0 \bigg(x_{12}
\, \sqrt{m^2+ q^2\, \alpha(1-z)} \bigg)\,,\label{aux.fun2}\\
F_0 (\b x_1, \b x_2;z)&= \int_0^\infty d q \,  q \, J_0( q  u) \, K_0 \bigg(x_{12}
\, \sqrt{m^2+ q^2\, z(1-z)} \bigg)\,,\label{aux.fun3}
\end{align}
where  $\b q = \b k_1 + \b k_2$. Substituting into \eq{kernel}  we derive 
\begin{align}\label{phi11}
\Phi_{11} (\b x_1, \b x_2; \b x_1' ,\b x_2'  ;  z )  =  4  C_F  
\bigg(\frac{\as}{\pi}\bigg)^2  \bigg\{ F_2(\b x_1, \b x_2;  z ) \, 
F_2 (\b x_1' , \b x_2' ;  z ) \, \frac{1}{x_{12} x_{12}'  u  u'} 
[ (1 - 2  z )^2 &\nonumber\\
\times\,  (\b x_{12} \cdot \b u)  (\b x_{12}' \cdot \b u') 
+ (\epsilon_{ij} \, u_i \, x_{12 \, j}) \, (\epsilon_{kl}  u'_k  x_{12 \, l}')] +
F_1(\b x_1, \b x_2;  z ) \, F_1(\b x_1' , \b x_2' ;  z )  m^2 
\frac{\b u\cdot\b u'}{u u'} & \nonumber\\
+ 4  z ^2  (1- z )^2 \,  F_0(\b x_1, \b x_2;  z )  
F_0(\b x_1' , \b x_2' ;  z )  - 2   z   (1 -  z )  (1 - 2 z ) 
\bigg[ \frac{\b x_{12} \cdot \b u}{x_{12} \, u}  F_2(\b x_1, \b x_2;  z ) &\nonumber\\
\times \,
F_0(\b x_1' , \b x_2' ;  z ) +  
\frac{\b x_{12}' \cdot \b u'}{x_{12}'  u'} 
F_2(\b x_1' , \b x_2' ;  z ) F_0(\b x_1, \b x_2;  z ) \bigg] \, \bigg\} \,,&
\end{align}
\begin{align}\label{phi22}
\Phi_{22} (\b x_1, \b x_2; \b x_1' ,\b x_2' ;  z )  =  4  C_F 
\bigg(\frac{\as}{\pi}\bigg)^2  m^2 \bigg\{
K_1(m x_{12})  K_1(m  x_{12}') \frac{1}{x_{12}  x_{12}'  u^2  u'^2} 
 [ (1 - 2   z )^2&\nonumber\\
\times\,(\b x_{12} \cdot \b u)  (\b x_{12}' \cdot \b u') 
+ (\epsilon_{ij} \, u_i \, x_{12 \, j})  (\epsilon_{kl} \, u'_k \, x_{12 \, l}')] +
K_0(m  x_{12}) K_0(m  x_{12}')  \frac{\b u\cdot\b u'}{u^2 \, u'^{2}} \bigg\}\,,&
\end{align}
\begin{align}\label{phi12}
\Phi_{12} (\b x_1, \b x_2; \b x_1' , \b x_2' ;  z ) = -  4  C_F  
\bigg(\frac{\as}{\pi}\bigg)^2  m 
\bigg\{ F_2 (\b x_1, \b x_2; z ) \, K_1(m  x_{12}')  
\frac{1}{x_{12}  x_{12}'  u  u'^2}  [ (1 - 2   z )^2
&\nonumber\\
\times\, (\b x_{12} \cdot \b u) \, (\b x_{12}' \cdot \b u') 
+ (\epsilon_{ij} \, u_i \, x_{12 \, j})  (\epsilon_{kl} \, u'_k \, x_{12 \, l}')]
+ m  F_1(\b x_1,\b x_2;  z ) 
 K_0(m  x_{12}')  \frac{\b u\cdot\b u'}{u u'^{2}} 
&\nonumber\\
- 2   z   (1 -  z ) (1 - 2   z ) 
\frac{\b x_{12}' \cdot \b u'}{x_{12}'  u'^2}   
F_0(\b x_1, \b x_2;  z )  K_1 (m  x_{12}') \bigg\}\,, 
\end{align}
All other products of the wave functions can be found using relations
\begin{align}
\Phi_{33} (\b x_1, \b x_2; \b x_1' , \b x_2' ;  z ) =& \,
\Phi_{11} (\b x_1, \b x_2; \b x_1' , \b x_2' ;  z ) + 
\Phi_{22} (\b x_1, \b x_2; \b x_1' , \b x_2' ;  z ) +
\Phi_{12} (\b x_1, \b x_2; \b x_1' , \b x_2' ;  z ) 
\nonumber\\
&+ \Phi_{21} (\b x_1, \b x_2; \b x_1' , \b x_2' ;  z )\label{phi33}\\
\Phi_{13} (\b x_1, \b x_2; \b x_1' , \b x_2' ;  z ) \, =&
- \Phi_{11} (\b x_1, \b x_2; \b x_1' , \b x_2' ;  z ) 
- \Phi_{12} (\b x_1, \b x_2; \b x_1' , \b x_2' ;  z ) \label{phi13}\\
\Phi_{23} (\b x_1, \b x_2; \b x_1' , \b x_2' ;  z ) \, =&
- \Phi_{21} (\b x_1, \b x_2; \b x_1' , \b x_2' ;  z ) 
- \Phi_{22} (\b x_1, \b x_2; \b x_1' , \b x_2' ;  z ) \label{phi23}
\end{align}
and 
\beq\label{phi_symm}
\Phi_{ij} (\b x_1, \b x_2; \b x_1' , \b x_2' ;  z ) \, =\, 
\Phi_{ji}^* (\b x_1' , \b x_2' ; \b x_1, \b x_2;  z ).
\eeq
Here Eqs. (\ref{phi33}),(\ref{phi13}),(\ref{phi23}) follow from
\eq{ms3}. Eq.~\eq{phi_symm} allows one to obtain $\Phi_{21}$,  
$\Phi_{31}$ and $\Phi_{32}$ from \eq{phi12}, \eq{phi13} and
\eq{phi23}.

Rescattering of $q_v$, $q_v \, g$ and $q_v\, q\, \bar q$
configurations on a large nucleus brings in different factors, which
we label by $S_{ij}$. For the case of single-quark inclusive production 
cross section they were calculated in \cite{Tuchin:2004rb}, while  in the case of
double-inclusive production in \cite{JalilianMarian:2004da,Dominguez:2011wm}. 
These factors can be written down as a combination of color dipole and color quadrupole scattering amplitudes. For notational simplicity  we will assume the large-$N_c$ limit in which only color dipoles survive. Generalization of our results beyond the large $N_c$ limit is straightforward, though very bulky.

$S_{ij}$'s are conventionally expressed in terms of the gluon saturation momentum defined as
\beq
Q_s^2 = 4  \pi  \as^2  \rho \, T(\b b)
\eeq
where $\rho$ is the nucleon number density in the nucleus and $T(\b b)$  the nuclear profile function. The scattering factors (which are proportional to the two-point correlations functions)  are given by
\begin{align}\label{s-matrix}
&S_F(\b x) = e^{-\frac{1}{8}\b x^2\ln(1/x\Lambda)Q_s^2}\,, 
&S_A(\b x) = e^{-\frac{1}{4}\b x^2\ln(1/x\Lambda)Q_s^2}\,
\end{align}
for quark and gluon color dipoles correspondingly. Using these definitions we have \cite{Kovchegov:2006qn}
\begin{align}
&S_{11} (\b x_1, \b x_2; \b x_1' , \b x_2' ;  z ) =S_F(\b x_1-\b x_1')S_F(\b x_2-\b x_2')\,,\label{xi}\\
&S_{22} (\b x_1, \b x_2; \b x_1' , \b x_2' ;  z ) = S_A(\b u-\b u')\,,\label{xi2}\\
&S_{33} (\b x_1, \b x_2; \b x_1' , \b x_2' ;  z ) = 1\,,\label{xi3}\\
&S_{12} (\b x_1, \b x_2; \b x_1' , \b x_2' ;  z ) =S_F(\b x_1-\b u') S_F(\b x_2-\b u')\,,\label{xi4}\\
&S_{23} (\b x_1, \b x_2; \b x_1' , \b x_2' ;  z ) = S_A(\b u)\,,\label{xi5}\\
&S_{13} (\b x_1, \b x_2; \b x_1' , \b x_2' ;  z ) = S_F(\b x_1)S_F(\b x_2)\,,\label{xi6}
\end{align}
where $\Lambda$ is an infrared cutoff. All other $S_{ij}$'s can be
found from the components listed in
\eq{xi} using
\beq\label{xi7}
S_{ij} (\b x_1, \b x_2; \b x_1' , \b x_2' ;  z ) =  
S_{ji} (\b x_1' , \b x_2' ; \b x_1, \b x_2;  z )
\eeq
similar to \eq{phi_symm}.

\section{Limit of proton collinear factorization}\label{sec:limit}

\subsection{Gluon scattering on heavy nucleus}\label{sec:gs}

\begin{figure}[ht]
      \includegraphics[height=3cm]{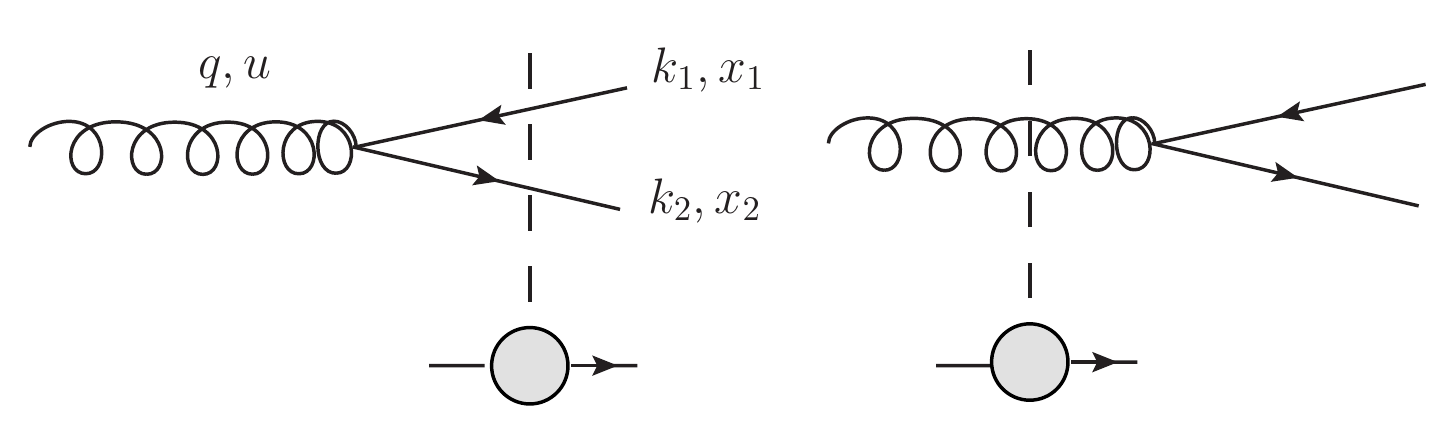} 
  \caption{Gluon-nucleus interactions contribution in the proton collinear approximation. }
\label{figX}
\end{figure}
Suppose now that valence quarks do not contribute to the quark-antiquark production. In this case the relevant diagrams in the light-cone perturbation theory are depicted in \fig{figX}.
The corresponding ``wave-functions" in the two cases differ only by the relative sign and are given by 
\begin{align}\label{za1}
\psi^{g\to q\bar q}_{\lambda,\sigma,\sigma'}(\b k_1, \b k_2) =\frac{gT^a}{k_1^-+k_2^- - q^-}\frac{\bar u_\sigma(k_1)}{\sqrt{k_1^+}}\gamma\cdot \epsilon^\lambda\frac{v_{\sigma'}(k_2)}{\sqrt{k_2^+}} \,.
\end{align}
Since integration over $\b q$ yields $(2\pi)^2\delta(\b u)$ we can consider the light-cone wave functions in the reference frame where  $\b q=0$ and integrate only over the remaining momentum $\b q'= \b k_2= -\b k_1$. 
\begin{align}\label{za2}
\psi^{g\to q\bar q}_{\lambda,\sigma,\sigma'}(\b k_1, \b k_2)= &-\frac{gT^a}{(1-z)\b k_1^2+z\b k_2^2+m^2}\nonumber\\
&\times\left\{\b \epsilon^\lambda\cdot [(1-z)\b k_1-z\b k_2](1-2z+\lambda\sigma)\delta_{\sigma,\sigma'}
+\frac{1}{\sqrt{2}}\sigma m(1-\lambda\sigma)\delta_{\sigma,-\sigma'}\right\}
\end{align} 
It is convenient to change variables from $\b x_{1,2}$ and $\b x_{1,2}'$ to $\b u$, $\b u'$, $\b x_{12}$ and $\b x_{12}'$:
\begin{align}
&\b x_1  = \b u +(1-z)\,\b x_{12}  & \b x_1' =\b u'+(1-z)\,\b x_{12}\label{w1}'\\
&\b x_2  = \b u -z\,\b x_{12}  & \b x_2' =\b u'-z\,\b x_{12}'\,.
\label{w2}
\end{align}
Performing Fourier transformation yields
\begin{align}\label{za3}
\psi^{g\to q\bar q}_{\lambda,\sigma,\sigma'}(\b x_1, \b x_2)= \frac{g T^a}{2\pi}\delta(\b u) \bigg\{
imK_1(mx_{12})\frac{\b \epsilon^\lambda\cdot \b x_{12}}{x_{12}}(1-2\alpha+\lambda\sigma)\delta_{\sigma,\sigma'}&\nonumber\\
-\delta_{\sigma,-\sigma}\frac{1}{\sqrt{2}}\sigma m (1-\lambda \sigma)K_0(mx_{12})\bigg\}\,.&
\end{align}
Multiplying by the contribution of the c.c.\ diagram, summing over the quantum numbers of the final states and averaging over the quantum numbers of initial particles according to 
\begin{align}\label{za4}
\phi^{q\to q\bar q}(\b x_{12}, \b x_{12}', z)= \frac{1}{2N_c^2}\Tr \sum_{\sigma,\lambda} \psi^{q\to q\bar q}_{\lambda,\sigma,\sigma'}(\b x_1, \b x_2)\psi^{q\to q\bar q *}_{\lambda,\sigma,\sigma'}(\b x_{1'}, \b x_{2}')
\end{align}
we derive
\begin{align}\label{phi.qq}
\phi^{q\to q\bar q}(\b x_{12}, \b x_{12}', z)=\frac{g^2}{(2\pi)^2}\delta(\b u)\delta(\b u')m^2\bigg\{
\frac{\b x_{12}\cdot \b x_{12}'}{x_{12} x_{12}'} [(1-z)^2+z^2] K_1(m x_{12})
K_1(m x_{12}')&\nonumber\\
 +K_0(m x_{12}) K_0(m x_{12}')\bigg\}\,.&
\end{align}
In particular, in the chiral limit \eq{phi.qq} reduces to 
\begin{align}\label{phi.qq.ch}
\phi^{g\to q\bar q}(\b x_{12}, \b x_{12}', z)=\frac{2g^2}{(2\pi)^2}\delta(\b u)\delta (\b u')\frac{\b x_{12}\cdot \b x_{12}'}{x_{12}^2 x_{12}'^2} P_{qg}(z)\,,
\end{align}
where 
\beql{split.qq}
P_{qg}(z) = \frac{1}{2}[(1-z)^2+ z^2]
\eeq
is the the splitting function. 

The cross section reads
\begin{align}\label{pA-qq}
\frac{d\sigma^{g\to q\bar qX}}{d^2k_1 d^2k_2\,dy\, dz}= \frac{1}{(2\pi)^4}\varphi(x_p,q^2)\frac{1}{\mathcal{S}_\bot} \int d^2x_1 d^2x_2 d^2 x_1'd^2x_2'd^2u\, d^2u'\, e^{-i\b k_1\cdot (\b x_1-\b x_1')-i\b k_2\cdot (\b x_2-\b x_2')}&\nonumber\\
 \times \phi^{g\to q\bar q}(\b x_{12}, \b x_{12}', z)\bigg[ S_F(\b x_1-\b x_1')S_F(\b x_2-\b x_2') -
S_F(\b x_1')S_F(\b x_2') -
S_F(\b x_{1})S_F(\b x_{2})  +1\bigg]\,,&
\end{align}
where $\mathcal{S}_\bot$ is the transverse cross-sectional area of nucleus. Integrals over $\b u$ and $\b u'$ are trivial due to the delta-functions in \eq{phi.qq.ch}. It is convenient to express the final result in terms of the transverse moments $\b q$ and $\b \ell$ defined as  
\begin{align}\label{dd1}
&\b q =\b k_1+\b k_2 \,, & \b \ell = (1-z)\b k_1-z\b k_2\,.
\end{align}
$\b q$ is the gluon transverse momentum, i.e.\ total momentum of the quark and antiquark. $\ell$ is the relative transverse momentum of the pair. Invariant mass of the pair $M$ can be written in terms of $\ell$ as follows
\beql{inv.mass}
M^2=(k_1+k_2)^2=\frac{m^2+\b\ell^2}{z(1-z)}\,.
\eeq
Substituting \eq{w1} and \eq{w2} and performing integration over the impact parameter $\b b'= (\b x_1'+\b x_2')/2$, which yields $\mathcal{S}_\bot$, we get
\begin{align}\label{pA-qq-2}
\frac{d\sigma^{g\to q\bar qX}}{d^2\ell\, d^2q\,dy\, dz\, d^2b}= &\frac{1}{(2\pi)^4}\frac{\as}{\pi}\varphi(x_p,q^2)\int d^2x_{12} \, d^2x_{12}' \, e^{-i\b \ell\cdot (\b x_{12}-\b x_{12}')}\nonumber\\
&\times\bigg\{ K_1(mx_{12})K_1(mx_{12}')\frac{\b x_{12} \cdot \b x_{12}' }{x_{12}  x_{12}' } [(1-z)^2+z^2]+ K_0(mx_{12})K_0(mx_{12}')\bigg\}\nonumber\\
& \times\bigg\{ S_F\big((1-z)(\b x_{12}-\b x_{12}')\big)S_F\big(z(\b x_{12}-\b x_{12}')\big) 
\nonumber\\
&-
S_F\big((1-z)\b x_{12}'\big)S_F(z\b x_{12}') -S_F\big((1-z)\b x_{12}\big)S_F(z\b x_{12})  +1\bigg\}\,.
\end{align}
This is the formula derived before in \cite{Kopeliovich:2002yv,Tuchin:2004rb}.

\subsection{Approximation of the general result of \sec{sec:vs}}\label{sec:der}

Now we would like to find an approximation to the general formulas of \sec{sec:vs} that lead to the same result \eq{pA-qq-2}. Collinear factorization of proton involves several assumptions. First,  the characteristic transverse momentum scale of proton $\Lambda$  is assumed to be much smaller than the produced quark mass. As a consequence, the size of color dipoles $\b u$ and $\b u'$ is much larger than that of $\b x_{12}$ and $\b x_{12}'$ i.e.\  
\beql{appr1}
x_{12}\ll u\ll \Lambda^{-1}\,. 
\eeq
Second, factorization requires that gluon coordinate be  the same in the amplitude and in the c.c.\ one independently of the produced quark and antiquark coordinates. This is ensured if
\beql{appr2}
|\b u- \b u'|\ll Q_s^{-1}\,. 
\eeq
Taking approximation \eq{appr1} and \eq{appr2} in \eq{xi} leads to the following expressions of the scattering matrix elements:
\begin{align}
&S_{11}(\b x_1, \b x_2; \b x_1' , \b x_2' ;  z ) 
\approx  S_F\big((1-z)(\b x_{12}-\b x_{12}')\big)S_F\big(z(\b x_{12}-\b x_{12}')\big)\,,\label{xii1}\\
&S_{22} (\b x_1, \b x_2; \b x_1' , \b x_2' ;  z ) \approx 1\,,\label{xii2}\\
&S_{33}(\b x_1, \b x_2; \b x_1' , \b x_2' ;  z ) =1\,,\label{xii3}\\
&S_{12}(\b x_1, \b x_2; \b x_1' , \b x_2' ;  z )   \approx S_F\big((1-z)\b x_{12}\big)S_F(z\b x_{12})\,,\label{xii4}\\
&S_{23}(\b x_1, \b x_2; \b x_1' , \b x_2' ;  z )  =S_A(\b u)\,,\label{xii5}\\
&S_{13}(\b x_1, \b x_2; \b x_1' , \b x_2' ;  z ) \approx  S_A(\b u)\,.\label{xii6}
\end{align}
Double inclusive cross section \eq{dcl} can be written as 
\begin{align}\label{dcl2}
\frac{d  \sigma^{q_v\to q\bar q X}}{d^2k_1  d^2 k_2  dy \, d  z  \, d^2 b} =& 
\frac{1}{4  (2  \pi)^6}  \int d^2 u\,  d^2 u'  d^2 x_{12}  d^2 x_{12}' \, 
e^{- i  (\b k_1 +\b k_2)\cdot (\b u - \b u')}\, e^{-i [ (1-z)\b k_1- z\b k_2]\cdot (\b x_{12}-\b x_{12}') } 
\nonumber\\
&\times \,  
\sum_{i,j=1}^3\, \Phi_{ij} (\b x_1, \b x_2; \b x_1' , \b x_2' ;  z ) \
S_{ij} (\b x_1, \b x_2; \b x_1' , \b x_2' ;  z )\,.
\end{align}
To take integrals over $\b u$ and $\b u'$ we note that  that $S_{ij}$, with $i,j\neq 3$ do not depend on these coordinate use the following identities  
\begin{align}
&\int  e^{-i \b q\cdot \b u}F_2 (\b x_1, \b x_2;z)\frac{\b u}{u}\, d^2u=-2\pi i\frac{\b q}{q^2}
K_1(x_{12}\sqrt{m^2+z(1-z)q^2})\sqrt{m^2+z(1-z)q^2} \label{f1}
 \,,\\
& \int  e^{-i \b q\cdot \b u}F_1 (\b x_1, \b x_2;z)\frac{\b u}{u}\, d^2u=-2\pi i\frac{\b q}{q^2}K_0(x_{12}\sqrt{m^2+z(1-z)q^2}) \, \label{f2},\\
& 
\int  e^{-i \b q\cdot \b u}F_0 (\b x_1, \b x_2;z)d^2u=2\pi K_0(x_{12}\sqrt{m^2+z(1-z)q^2})\,,
\label{f3}\\
&
\int e^{-\b q\cdot \b u}\, \frac{\b u}{u^2}\, d^2u = -2\pi i \frac{\b q}{q^2}\label{f4}\,.
\end{align}
In derivation of these identities we used the orthogonality of the Bessel functions
\beq\label{f5}
\int_0^\infty  u\, J_\alpha(q u)J_\alpha(q' u) \, du= \frac{1}{q}\,\delta (q-q')\,.
\eeq

Denoting $s= \sqrt{m^2+z(1-z)q^2}$ we write 
\begin{align}\label{aa2}
&\int d^2 u \int d^2 u' \,\Phi_{11} (\b x_1, \b x_2; \b x_1' ,\b x_2'  ;  z ) e^{-i\b q \cdot \b u +i \b q\cdot \b u'} 
 \nonumber\\
&
=  4  C_F 
\bigg(\frac{\as}{\pi}\bigg)^2 \, (2\pi)^2 \,\bigg\{ K_1(x_{12}s )K_1(x_{12}'s)s^2 
 \frac{1}{x_{12}  x_{12}'  q^4} [ (1 - 2   z )^2  (\b x_{12} \cdot \b q)  (\b x_{12}' \cdot \b q) 
\nonumber\\
&
+ (\epsilon_{ij} q_i  x_{12 \, j}) \, (\epsilon_{kl}  q_k  x_{12 \, l}')] 
+K_0(x_{12}s)  K_0(x_{12}'s) \frac{m^2}{q^2} 
+ 4   z ^2 (1- z )^2  K_0(x_{12}s) 
K_0(x_{12}'s ) 
\nonumber\\
&
 - 2   z   (1 -  z )  (1 - 2  z )
\bigg[ \frac{\b x_{12} \cdot \b q}{x_{12} q^2} (-i)\, K_1(x_{12}s)K_0(x_{12}'s)s
+  
\frac{\b x_{12}' \cdot \b q}{x_{12}' q^2}i \, 
K_1(x_{12}'s)K_0(x_{12}s)s\bigg] \, \bigg\} \,,
\end{align}

\begin{align}\label{aa3}
&\int d^2 u \int d^2 u' \,\Phi_{22} (\b x_1, \b x_2; \b x_1' ,\b x_2'  ;  z ) e^{-i\b q \cdot \b u +i \b q\cdot \b u'} 
 \nonumber\\
&
=  4  C_F 
\bigg(\frac{\as}{\pi}\bigg)^2 (2\pi)^2 
m^2\bigg\{ K_1(x_{12}m )K_1(x_{12}'m)
 \frac{1}{x_{12}  x_{12}'  q^4} [ (1 - 2   z )^2  (\b x_{12} \cdot \b q)  (\b x_{12}' \cdot \b q) 
\nonumber\\
&
+ (\epsilon_{ij} q_i  x_{12 \, j}) \, (\epsilon_{kl}  q_k  x_{12 \, l}')] 
+K_0(x_{12}m)  K_0(x_{12}'m) \frac{1}{q^2} \bigg\}\,,
\end{align}

\begin{align}\label{aa4}
&\int d^2 u \int d^2 u' \,\Phi_{12} (\b x_1, \b x_2; \b x_1' ,\b x_2'  ;  z ) e^{-i\b q \cdot \b u +i \b q\cdot \b u'} 
 \nonumber\\
&
=  -4  C_F 
\bigg(\frac{\as}{\pi}\bigg)^2 \, (2\pi)^2 \,m\bigg\{ K_1(x_{12}s )K_1(x_{12}'m)s 
 \frac{1}{x_{12}  x_{12}'  q^4} [ (1 - 2   z )^2  (\b x_{12} \cdot \b q)  (\b x_{12}' \cdot \b q) 
\nonumber\\
&
+ (\epsilon_{ij} q_i  x_{12 \, j}) \, (\epsilon_{kl}  q_k  x_{12 \, l}')] 
+mK_0(x_{12}s)  K_0(x_{12}'m) \frac{1}{q^2} 
\nonumber\\
&
 - 2   z   (1 -  z )  (1 - 2  z ) 
\frac{\b x_{12}' \cdot \b q}{x_{12}' q^2}i \, 
K_1(x_{12}'m)K_0(x_{12}s)  \bigg\} \,.
\end{align}

In the factorization limit, only terms with $i,j=1,2$ contribute as we demonstrate in the next section. Therefore, in this section we concentrate only $i,j=1,2$ terms. The collinear limit comes from the  logarithmically enhanced terms  that correspond to the most singular in $1/q$ behavior. Thus, in order to obtain the collinear factorization of proton, we need to assume  that  
\beql{appr3}
q\ll 2m\,,
\eeq
which implies that $|\b u-\b u'|\gg m^{-1}$. In view of \eq{appr1} it means that we require $Q_s\ll m$.  Using \eq{appr3} in  Eqs.~\eq{aa2},\eq{aa3},\eq{aa4} we conclude that they coincide (apart from the relative sign):
 \begin{align}\label{zz1}
&\int d^2 u \int d^2 u' \,\Phi_{11}\, e^{-i\b q \cdot \b u +i \b q\cdot \b u'} \approx \int d^2 u \int d^2 u' \,\Phi_{22} \, e^{-i\b q \cdot \b u +i \b q\cdot \b u'} \approx -
\int d^2 u \int d^2 u' \,\Phi_{12}\, e^{-i\b q \cdot \b u +i \b q\cdot \b u'} 
 \nonumber\\
&
\approx  4  C_F 
\bigg(\frac{\as}{\pi}\bigg)^2 \, (2\pi)^2 m^2 \,\bigg\{ K_1(x_{12}m )K_1(x_{12}'m)
 \frac{1}{x_{12}  x_{12}'  q^4} [ (1 - 2   z )^2  (\b x_{12} \cdot \b q)  (\b x_{12}' \cdot \b q) 
\nonumber\\
&
+ (\epsilon_{ij} q_i  x_{12 \, j}) \, (\epsilon_{kl}  q_k  x_{12 \, l}')] 
+K_0(x_{12}m)  K_0(x_{12}'m) \frac{1}{q^2} \bigg\}\,.
\end{align}
 Substituting \eq{zz1} and \eq{xii1}-\eq{xii6} into \eq{dcl2} and averaging over directions of $\b q $ using $\aver{q_iq_j}= q^2\delta_{ij}/2$ and $\epsilon_{ij}\epsilon_{il}=2\delta_{jl}$ we obtain
\begin{align}\label{dcl3}
\frac{d \sigma_\text{fact}^{q_v\to q\bar q X}}{d^2\ell \, d^2q\, dy \, d  z  \, d^2 b} =&
\frac{1}{  (2  \pi)^4}    C_F 
\bigg(\frac{\as}{\pi}\bigg)^2 \, \frac{m^2}{q^2}\, \int   d^2 x_{12}  d^2 x_{12}' \, 
 e^{-i \b\ell\cdot (\b x_{12}-\b x_{12}') } 
\nonumber\\
&\times \, \bigg\{ K_1(x_{12}m )K_1(x_{12}'m) 
 \frac{\b x_{12} \cdot \b x_{12}' }{x_{12}  x_{12}' } [ (1 -  z )^2+z^2]  
+K_0(x_{12}m)  K_0(x_{12}'m)   \bigg\} \nonumber\\
&\times \bigg\{ S_F\big((1-z)(\b x_{12}-\b x_{12}')\big)S_F\big(z(\b x_{12}-\b x_{12}')\big)
+1\nonumber\\
&-S_F\big((1-z)\b x_{12}\big)S_F(z\b x_{12})-S_F\big((1-z)\b x_{12}'\big)S_F(z\b x_{12}')\bigg\}\,.
\end{align}
Unintegrated gluon distribution function $\varphi(x_p,q)$ defined such that 
\beql{ugd}
x_pG(x_p,Q^2)=\int^{Q^2}\varphi(x_p,q^2) \, dq^2\,.
\eeq
Since in the leading logarithmic approximation  
\beq\label{xg}
xG(x,Q^2)= \frac{\as C_F}{\pi}\ln \frac{Q^2}{\Lambda^2}\,,
\eeq
we can write \eq{dcl3} as
\begin{align}\label{dcl4}
\frac{d \sigma_\text{fact}^{q_v\to q\bar q X}}{d^2\ell \, d^2q\, dy \, d  z  \, d^2 b} =&
\frac{1}{  (2  \pi)^4}    
\frac{\as}{\pi} \, m^2\,\varphi(x_p,q^2) \int   d^2 x_{12}  d^2 x_{12}' \, 
 e^{-i \b\ell\cdot (\b x_{12}-\b x_{12}') } 
\nonumber\\
&\times \, \bigg\{ K_1(x_{12}m )K_1(x_{12}'m) 
 \frac{\b x_{12} \cdot \b x_{12}' }{x_{12}  x_{12}' } [ (1 -  z )^2+z^2]  
+K_0(x_{12}m)  K_0(x_{12}'m)   \bigg\} \nonumber\\
&\times \bigg\{ S_F\big((1-z)(\b x_{12}-\b x_{12}')\big)S_F\big(z(\b x_{12}-\b x_{12}')\big)
+1\nonumber\\
&-S_F\big((1-z)\b x_{12}\big)S_F(z\b x_{12})-S_F\big((1-z)\b x_{12}'\big)S_F(z\b x_{12}')\bigg\}\,.
\end{align} 
As expected this coincides with \eq{pA-qq-2}.

If we are interested in single inclusive cross section, we have to integrate \eq{dcl4} over $\b q$  and multiply by 2, which yields
\begin{align}\label{id4}
\frac{d\sigma_\text{fact}^{q_v\to q X}}{d^2\ell\, dy\, d^2b}= &\frac{\as}{8\pi^4}\,m^2\int_0^1 dz \int d^2 x_{12}\int d^2 x_{12}'\,xG(x,1/|\b x_{12}-\b x_{12}'|)\, 
e^{-i\b\ell\cdot (\b x_{12}-\b x_{12}')}\nonumber\\
&\times\bigg\{ K_1(mx_{12})K_1(mx_{12}')\frac{\b x_{12} \cdot \b x_{12}' }{x_{12}  x_{12}' } [(1-z)^2+z^2]+ K_0(mx_{12})K_0(mx_{12}')\bigg\}\nonumber\\
&\times \bigg\{ S_F\big((1-z)(\b x_{12}-\b x_{12}')\big)S_F\big(z(\b x_{12}-\b x_{12}')\big)
+1\nonumber\\
&-S_F\big((1-z)\b x_{12}\big)S_F(z\b x_{12})-S_F\big((1-z)\b x_{12}'\big)S_F(z\b x_{12}')\bigg\}\,.
\end{align}

\section{Contribution of valence quarks}\label{sec:ccc}

In the factorization approximation valence quarks do not contribute. Indeed, this is evident upon substitution  \eq{zz1} into \eq{phi33}--\eq{phi23}: $\Phi_{13}=\Phi_{23}=\Phi_{33}=0$.  Contribution of valence quarks emerges, along with other contributions, when we keep the logarithmically sub-leading terms in \eq{aa2}-\eq{aa4}. In other words, we are still going to work within the approximations \eq{appr1},\eq{appr2}, but relax the approximation \eq{appr3}. Thus, contribution of valence quarks arises when $Q_s>m$. 

It is convenient to rewrite the last line in \eq{dcl}  using \eq{phi33}--\eq{phi23} as (dropping the summation sign)
\begin{align}\label{wp1}
&\Phi_{ij}S_{ij} =\left(\Phi_{ij}S_{ij}\right)_I+\left(\Phi_{ij}S_{ij}\right)_{II}
\end{align}
where
\begin{align}\label{wp1a}
 \left(\Phi_{ij}S_{ij}\right)_\text{I}=&\Phi_{11}S_{11}+\Phi_{12}S_{12}+\Phi_{21}S_{21}+\Phi_{22}S_{22}\,,\nonumber\\
 \left(\Phi_{ij}S_{ij}\right)_\text{II}=
&\Phi_{11}(S_{33}-S_{23}-S_{31})+\Phi_{12}(S_{33}-S_{13}-S_{32})\nonumber\\
&+\Phi_{21}(S_{33}-S_{23}-S_{31})+\Phi_{22}(S_{33}-S_{32}-S_{23})
\,.
\end{align}
Because of  \eq{appr1}, $\left(\Phi_{ij}S_{ij}\right)_\text{I}$ depends only on $\b x_{12}$ and $\b x_{12}'$, whereas $\left(\Phi_{ij}S_{ij}\right)_\text{II}$ depends only on $\b u$ and $\b u'$. We will denote the corresponding contributions to the cross section by as $\sigma_\text{I}$ and $\sigma_\text{II}$ correspondingly. We detailed the calculation of $\sigma_\text{I}$ in \sec{sec:der}. The double-inclusive cross section averaged over the directions of $\b q$  is  
\begin{align}\label{wp2}
&\frac{d \sigma_\text{I}^{q_v\to q\bar q X}}{\pi dq^2\, d^2\ell \, dy \, d  z  \, d^2 b} =
\frac{1}{  (2  \pi)^4}\,    C_F 
\bigg(\frac{\as}{\pi}\bigg)^2\, \int   d^2 x_{12}\,  d^2 x_{12}' \, 
 e^{-i \b\ell\cdot (\b x_{12}-\b x_{12}') }
\nonumber\\
&\times \, \bigg\{
 \frac{\b x_{12} \cdot \b x_{12}' }{x_{12}  x_{12}' } [ (1 -  z )^2+z^2]  \,\frac{1}{q^2}
 \big [K_1(x_{12}s)K_1(x_{12}'s)s^2 S_{11} + K_1(x_{12}m)K_1(x_{12}'m)m^2 S_{22}\nonumber\\
&- K_1(x_{12}s)K_1(x_{12}'m)ms S_{12}-K_1(x_{12}m)K_1(x_{12}'s)ms S_{21}\big] 
+\frac{m^2}{q^2}\big[K_0(x_{12}s)  K_0(x_{12}'s)S_{11}\nonumber\\
&+K_0(x_{12}m)  K_0(x_{12}'m)S_{22}-K_0(x_{12}s)  K_0(x_{12}'m)S_{12}
-K_0(x_{12}m)  K_0(x_{12}'s)S_{21}\big ]
 \nonumber\\
&+ 4z^2(1-z)^2K_0(x_{12}s)K_0(x_{12}'s)S_{11} \bigg\}\,.
\end{align}
Integral over $q$ is logarithmically divergent in the UV, with the cutoff $Q$ such that $q\le Q$.

Now we turn to the contribution $\sigma_\text{II}$. 
Substituting \eq{xii1}--\eq{xii6} we get
\begin{align}\label{wx7}
\frac{d\sigma_\text{II}^{q_v\to q\bar qX}}{d^2k_1 d^2k_2dy\, d^2b\,dz}= &\frac{1}{4(2\pi)^6}\int d^2u\,d^2u' \, e^{-i\b q\cdot(\b u-\b u')}\int d^2x_{12}\, d^2x_{12}'\, e^{-i\b \ell\cdot (\b x_{12}-\b x_{12}')}\,\nonumber\\
&
\times\, [1-S_A(\b u)-S_A(\b u')]\,\big\{ \Phi_{11}(\b x_1, \b x_2; \b x_1' ,\b x_2'  ;  z ) + \Phi_{12}(\b x_1, \b x_2; \b x_1' ,\b x_2'  ;  z )\nonumber\\
&  +\Phi_{21}(\b x_1, \b x_2; \b x_1' ,\b x_2'; z )+\Phi_{22}(\b x_1, \b x_2; \b x_1' ,\b x_2'  ;  z ) \big\}\,.
\end{align}
Since the scattering matrix elements are independent of $\b x_{12}$ and $\b x_{12}'$, we can integrate the wave function  over these variables. The integrations are performed employing the following identities
\begin{align}\label{wx1}
&\int d^2x_{12} e^{-i\b \ell \cdot \b x_{12}}F_2(\b x_1, \b x_2, z)\frac{\b x_{12}}{x_{12}}= 
-2\pi i \b \ell \int_0^\infty dq \,J_1(qu) \frac{1}{\ell^2+z(1-z)q^2+m^2}\nonumber\\
&
=-\frac{2\pi i \b \ell}{(\ell^2+m^2)u}\left[ 1- u\sqrt{\frac{\ell^2+m^2}{z(1-z)}}\, K_1\left( u 
\sqrt{\frac{\ell^2+m^2}{z(1-z)}}\right)\right]\equiv -\frac{2\pi i \b \ell}{(\ell^2+m^2)u}f_1( u, \ell, z)\,.
\end{align}

\begin{align}\label{wx2}
&\int d^2x_{12} \, e^{-i\b \ell \cdot \b x_{12}}F_1(\b x_1, \b x_2, z)= 
2\pi  \int_0^\infty dq \,J_1(qu) \frac{1}{\ell^2+z(1-z)q^2+m^2}\nonumber\\
&= \frac{2\pi}{(\ell^2+m^2)u}f_1( u, \ell, z)\,.
\end{align}

\begin{align}\label{wx3}
&\int d^2x_{12} e^{-i\b \ell \cdot \b x_{12}}F_0(\b x_1, \b x_2, z)= 
2\pi  \int_0^\infty dq \,J_0(qu) \frac{1}{\ell^2+z(1-z)q^2+m^2}\nonumber\\
&
=\frac{2\pi }{z(1-z)} K_0\left( u 
\sqrt{\frac{\ell^2+m^2}{z(1-z)}}\right)\equiv \frac{2\pi }{z(1-z)} f_0( u,  \ell, z)\,.
\end{align}

Using these formulas we have
\begin{align}\label{wx4}
&\int d^2x_{12}\int d^2 x_{12}' \,e^{-i\b \ell\cdot \b x_{12}+i\b\ell \cdot \b x_{12}'}\,\Phi_{11}(\b x_1, \b x_2;\b x_1',\b x_2';z)=\nonumber\\
&=4C_F \left( \frac{\as}{\pi}\right)^2(2\pi)^2\bigg\{ \frac{1}{u^2u'^2}\frac{1}{(\ell^2+m^2)^2}[(1-2z)^2(\b u\cdot \b \ell)(\b u'\cdot \b \ell)+(\epsilon_{ij}u_i\ell_j)(\epsilon_{kl}u'_k\ell_l)]
\nonumber\\
&\times\, f_1(u,\ell,z)f_1(u',\ell,z)+
 m^2\frac{\b u\cdot \b u'}{u^2u'^2}\frac{1}{(\ell^2+m^2)^2}f_1(u,\ell,z)f_1(u',\ell,z)\nonumber\\
 &+4 z^2(1-z)^2\frac{1}{z^2(1-z)^2}f_0(u,\ell,z)f_0(u',\ell,z)\nonumber\\
&
 -2z(1-z)(1-2z)\bigg[ \frac{\b u\cdot \b \ell}{u^2}(-i)\frac{1}{z(1-z)}\frac{1}{\ell^2+m^2}f_1(u,\ell,z)f_0(u',\ell,z)\nonumber\\
&+\frac{\b u'\cdot \b \ell}{u'^2}i\frac{1}{z(1-z)}\frac{1}{\ell^2+m^2}f_0(u,\ell,z)f_1(u',\ell,z)\bigg]\bigg\}
\end{align}

\begin{align}\label{wx5}
&\int d^2x_{12}\int d^2 x_{12}' e^{-i\b \ell\cdot \b x_{12}+i\b\ell \cdot \b x_{12}'}\,\Phi_{22}(\b x_1, \b x_2;\b x_1',\b x_2';z)=\nonumber\\
&=4C_F \left( \frac{\as}{\pi}\right)^2(2\pi)^2m^2\bigg\{ \frac{1}{m^2u^2u'^2}\frac{1}{(\ell^2+m^2)^2}[(1-2z)^2(\b u\cdot \b \ell)(\b u'\cdot \b \ell)+(\epsilon_{ij}u_i\ell_j)(\epsilon_{kl}u'_k\ell_l)]
\nonumber\\
&+
 \frac{\b u\cdot \b u'}{u^2u'^2}\frac{1}{(\ell^2+m^2)^2}\bigg\}
\end{align}

\begin{align}\label{wx6}
&\int d^2x_{12}\int d^2 x_{12}' e^{-i\b \ell\cdot \b x_{12}+i\b\ell \cdot \b x_{12}'}\,\Phi_{12}(\b x_1, \b x_2;\b x_1',\b x_2';z)=\nonumber\\
&=-4C_F \left( \frac{\as}{\pi}\right)^2(2\pi)^2m\bigg\{ \frac{1}{mu^2u'^2}\frac{1}{(\ell^2+m^2)^2}[(1-2z)^2(\b u\cdot \b \ell)(\b u'\cdot \b \ell)+(\epsilon_{ij}u_i\ell_j)(\epsilon_{kl}u'_k\ell_l)]
\nonumber\\
&\times\, f_1(u,\ell,z)+
 m\frac{\b u\cdot \b u'}{u^2u'^2}\frac{1}{(\ell^2+m^2)^2}f_1(u,\ell,z)\nonumber\\
 &
 -2z(1-z)(1-2z) \frac{\b u'\cdot \b \ell}{u'^2}(-i)\frac{1}{z(1-z)}\frac{1}{\ell^2+m^2}f_0(u',\ell,z)\bigg\}
\end{align}
Fourier transforms of other wave functions can be derived using \eq{phi33}--\eq{phi23}. Substituting in \eq{wx7}, the resulting expression averaged over the directions of $\b \ell$ (to make it less bulky)  reads
\begin{align}\label{wx15}
\frac{d\sigma_\text{II}^{q_v\to q\bar qX}}{\pi d\ell^2\, d^2q\,dy d^2b\,dz}=&
\frac{1}{4(2\pi)^4}4C_F \left( \frac{\as}{\pi}\right)^2\int d^2u\int d^2u' \, e^{-i\b q\cdot (\b u-\b u')} \nonumber\\
&\times\bigg\{ \frac{\b u\cdot \b u'}{u^2u'^2}\frac{1}{(\ell^2+m^2)^2}[\ell^2((1-z)^2+z^2)+m^2](f_1(u,\ell,z)-1)(f_1(u',\ell,z)-1)\nonumber\\
&+ 4f_0(\b u,\ell,z)f_0(\b u',\ell,x)\bigg\}\,\big[1- S_A(\b u)-S_A(\b u')\big] \,.
\end{align}
To obtain the single inclusive cross section we can integrate over $\b\ell$ or over $\b q$ and multiple by 2 giving the final result
\begin{align}\label{wx18}
\frac{d\sigma_\text{II}^{q_v\to qX}}{d^2\ell dy d^2b}=&\frac{1}{2(2\pi)^2}\,4\,C_F \left( \frac{\as}{\pi}\right)^2\int _0^1 dz \int d^2u \,\big[1-2 S_A(\b u)\big]\nonumber\\
&\times \, \bigg\{\big[ \ell^2 ((1-z)^2+z^2) +m^2\big]\frac{(f_1(u,\ell,z)-1)^2}{u^2(\ell^2+m^2)^2} + 4f_0^2(u,\ell,z)\bigg\}\,.
\end{align}
Alternatively, we could have used \eq{wx7} to write
\begin{align}\label{wx9}
\frac{d\sigma_\text{II}^{q_v\to q\bar qX}}{d^2\ell\, dy d^2b\,dz}= &\frac{1}{2(2\pi)^4}\int d^2u\int d^2u' \,\delta(\b u-\b u')\, [1-2S_A(\b u)]\nonumber\\
&\times\bigg\{\int d^2x_{12}\, d^2x_{12}'\, e^{-i\b \ell\cdot (\b x_{12}-\b x_{12}')}
 [ \Phi_{11}(\b x_1, \b x_2; \b x_1' ,\b x_2'  ;  z ) +\Phi_{22}(\b x_1, \b x_2; \b x_1' ,\b x_2'  ;  z ) ] \nonumber\\
&+2\,\real\int d^2x_{12}\, d^2x_{12}'\, e^{-i\b \ell\cdot (\b x_{12}-\b x_{12}')}\Phi_{12}(\b x_1, \b x_2; \b x_1' ,\b x_2'  ;  z ) \bigg\}\,,
\end{align}
and plugg in formulas \eq{wx4}--\eq{wx6}. 

The final result, which is the main result of this paper is  
\begin{align}\label{main1}
\frac{d\sigma^{q_v\to q\bar qX}}{d^2k_1 d^2k_2dy\, d^2b\,dz}= \frac{d\sigma_\text{I}^{q_v\to q\bar qX}}{d^2k_1 d^2k_2dy\, d^2b\,dz}+\frac{d\sigma_\text{II}^{q_v\to q\bar qX}}{d^2k_1 d^2k_2dy\, d^2b\,dz}\,,
\end{align}
where the two terms on the right-hand-side are given by \eq{wp2} and \eq{wx15}.

Note that the separation between  the contributions $\sigma_\text{I}$ and $\sigma_\text{II}$ in \eq{main1}  is gauge dependent. Only the sum of the two cross section has the physical meaning as the heavy quark and antiquark production cross section. In the limit $u\ll 1/m$, it follows that $f_1\to 1$, $f_0\to 0$. If interactions are absent, i.e.\ $S_{ij}=1$ this implies that $\sigma_\text{II}\to 0$, so that $\sigma_\text{I}$ (given by \eq{id4}) acquires independent physical meaning (see the previous section). 

It can be explicitly shown that if we put $S_{ij}=1$ for all $i,j$ the cross section vanishes as required. Indeed, in this case integrating \eq{wp2} over $\b x_{12}$ and $\b x_{12}'$ and integrating \eq{wx15} over $\b u$ and $\b u'$ we arrive at 
\begin{align}\label{wx20}
&\frac{d\sigma_\text{I}^{q_v\to q\bar qX}}{d^2\ell\, d^2q\,dy d^2b\,dz}= -\frac{d\sigma_\text{II}^{q_v\to q\bar qX}}{d^2\ell\, d^2q\,dy d^2b\,dz}\nonumber\\
&=\frac{1}{(2\pi)^2}\left( \frac{\as}{\pi}\right)^2 C_F \frac{z^2(1-z)^2}{(z(1-z)q^2+\ell^2+m^2)^2}
\bigg\{ \frac{q^2}{(\ell^2+m^2)^2}\big[\big((1-z)^2+z^2\big)\ell^2+m^2\big]+4\bigg\}\,.
\end{align}
In order that the $q$-integrated cross section vanish when  $S_{ij}=1$, integral over $q$ in  \eq{wp2} must be cutoff by $q\le Q$, while  integral over $u$ in \eq{wx18}  by $u\ge 1/Q$ with the same $Q$.

\begin{figure}[ht]
      \includegraphics[height=5cm]{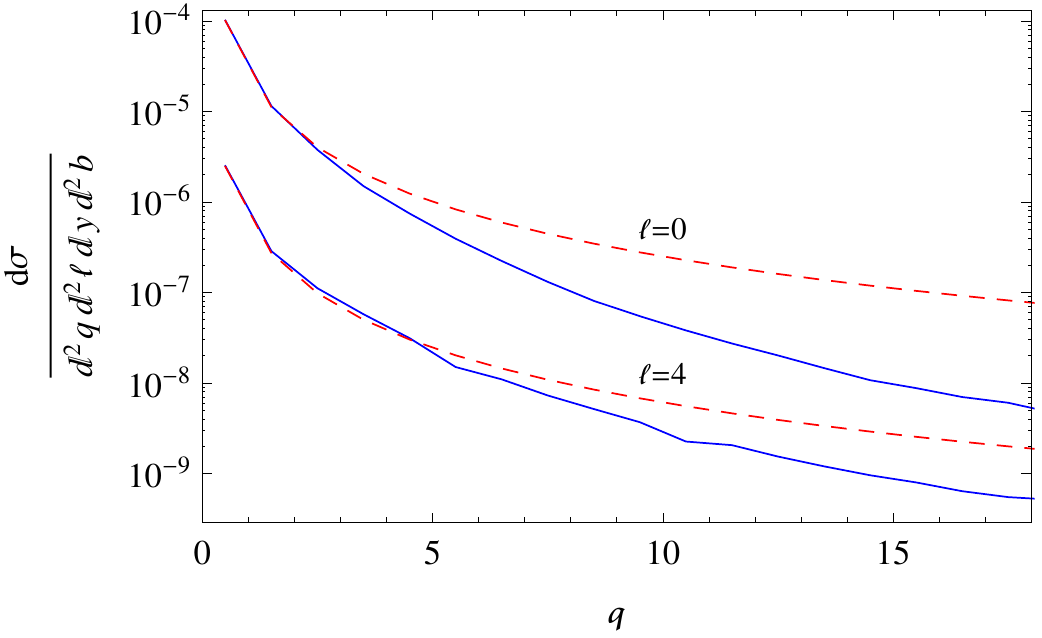} 
  \caption{Double inclusive cross section \eq{main1} for $m=1.29$~GeV (charm) and $Q_s=1$~GeV. Solid lines include contribution of valence quarks, while dashed lines correspond to the collinear factorization of proton. }
\label{fig3}
\end{figure}

The role of the valence quark interactions is illustrated in \fig{fig3}. It shows numerical calculation of charm production with and without valence quark contribution. One observes that collinear approximation fails at higher $q$ and lower $\ell$. 

\section{Conclusions }

We investigated the collinear factorization of the proton in the process of  heavy quark production in high energy pA collisions. The factorization holds in the regime $Q_s\ll m$, which also implies that the gluon saturation effects on the heavy nucleus side are small. However, if $Q_s\gtrsim m$ this approximation breaks down and one has to take into account interaction of valence quarks with the nucleus. Generally, the collinear factorization of proton is a reasonable approximation only at high relative transverse momentum $\b\ell$ of quark and antiquark (high invariant masses $M$), and small total transverse momentum $\b q$ of the pair  (gluon is collinear with the proton), i.e.\ for  hard scattering. These observations have important phenomenological implications that will be discussed in a separate work.

\acknowledgments
This work  was supported in part by the U.S. Department of Energy under Grant No.\ DE-FG02-87ER40371.


\end{document}